\documentclass{jfm}

\usepackage[usenames, dvipsnames]{color}
\usepackage{graphicx}
\usepackage{epstopdf, epsfig}
\usepackage{gensymb}
\usepackage{amsmath}
\usepackage{cleveref}
\usepackage{bm}
\usepackage{siunitx}
\graphicspath{{figures/}}

\begin{document}

\newtheorem{lemma}{Lemma}
\newtheorem{corollary}{Corollary}

\shorttitle{On the criteria of large cavitation bubbles} % for header on odd pages
\shortauthor{P. Xu, S. Liu, Z. Zuo  and Z. Pan} % for header on even pages. P. Xu et al.

\title{On the criteria of large cavitation bubbles in a tube during a transient process}

\author
 {
 Peng Xu\aff{1,2}, %Peng Xu
  Shuhong Liu\aff{1},%Shuhong Liu
    Zhigang Zuo\aff{1}%Zhigang Zuo
  \corresp{\email{zhigang200@tsinghua.edu.cn}},
  \and 
  Zhao Pan\aff{2}%Zhao Pan
  \corresp{\email{zhao.pan@uwaterloo.ca}}  }

\affiliation
{
\aff{1}
Department of Energy and Power Engineering, and State Key Laboratory of Hydro Science and Engineering, Tsinghua University, Beijing 100084, China. 
\aff{2}
Department of Mechanical and Mechatronics Engineering,  University of Waterloo, University of Waterloo
200 University Avenue West, 
Waterloo, Ontario N2L 3G1, Canada. 
}

\maketitle

\begin{abstract}  %%% --- ABSTRACT --- %%%
Extreme cavitation scenarios such as water column separations in hydraulic systems during transient processes caused by large cavitation bubbles can lead to catastrophic destruction. In the present paper we study the onset criteria and dynamics of large cavitation bubbles in a tube. A new cavitation number $Ca_2 = {l^*}^{-1} Ca_0$ is proposed to describe the maximum length $L_{\max}$ of the cavitation bubble, where $l^*$ is a non-dimensional length of the water column indicating its slenderness, and $Ca_0$ is the classic cavitation number.
% $Ca_0 = p_ \infty / 0.5 \rho u_0^2$ is the classic cavitation number, where $u_0$ is the characteristic velocity of the flow, and $\rho$ is the density of the liquid.
Combined with the onset criteria for acceleration-induced cavitation ($Ca_1<1$, \cite{Pan2017Cavitation}), we show that the occurrence of large cylindrical cavitation bubbles requires both $Ca_2<1$ and $Ca_1<1$ simultaneously.
We also establish a Rayleigh-type model for the dynamics of large cavitation bubbles in a tube. 
The bubbles collapse at a finite end speed, and the  time from the maximum bubble size to collapse is {\color{black} $T_c=\sqrt{2}\sqrt{lL_{\max}}\sqrt{\frac{\rho}{p_\infty}}$, where $l$ is the length of the water column, $L_{\max}$ is the maximum bubble length, $\rho$ is the liquid density, and $p_{\infty}$ is the reference pressure in the far field}. The analytical results are validated against systematic experiments using a modified `tube-arrest' apparatus, which can decouple acceleration and velocity. The results in the current work can guide design and operation of hydraulic systems encountering transient processes. 
\end{abstract}

%%% --- INTRODUCTION --- %%%
\section{Introduction}\label{intro} % * star eliminates numbering
Cavitation involves explosive growth of vapor bubbles subjected to a quick local pressure reduction, and subsequent destructive collapses. Examples of cavitation effects include sharp head drop of hydraulic machinery, erosion of spillways, and ultrasonic cleaning~\citep{inproceedings,  escaler2006detection, torre2011experimental}.
% ~\citep{li2000cavitation, inproceedings,  escaler2006detection, lu2017experimental, torre2011experimental}.
 Particularly, the formation and collapse of cavitation bubbles can induce large pressure pulsations and heavy loads on a hydraulic system such as pipelines and hydraulic turbines, etc.~\citep{benjamin1966collapse,avellan2004introduction,luo2016review}. 

An extreme case of cavitation is the formation of liquid column separation in a flow passage induced by intense transient processes. For example, when an upstream valve rapidly closes in a pipeline, the water downstream attempts to continue flowing, and generates local large cavitation bubbles occupying the entire cross-section. The abrupt collapses or implosions of these large cavitation bubbles in hydraulic pipelines can cause catastrophic destruction and result in casualties and substantial economic losses~\citep{bonin1960water,Bergant2006Water,Hillgren2011Analysis}. 

Previous studies have discussed the propagation and superposition of pressure oscillations induced by the collapse of the cavitation bubble and water hammer in simple pipeline systems. 
Experimental studies mainly focused on investigating the pressure variations in the pipelines and the validation of the relevant numerical simulations~\citep{Simpson1991,Bergant1999pipeline, sv-jmesv-jme.2014.1882,Adamkowski2015Cavitation}.
% {\color{magenta}In more complex systems, such as hydraulic turbines, numerical analyses on the internal flow and pressure field have been carried out during the close-up process of the guide vanes~\citep{Zhang2016CFD,Ke2019Runner}}.
In more complex systems, such as hydraulic turbines, {\color{black}the internal flow and pressure fields during guide vanes closing process have been numerically studied} ~\citep{Zhang2016CFD,Ke2019Runner}.
Their results suggest that a moderate piece-wise guide vane closing process, i.e., a mild flow deceleration, is beneficial to reduce the risk of cavitation during the transient process. %\zp{Why this is connected to cavitation? need one more sentence to build the connection.}
Through a water-hammer estimation, it has been recognized that the amplitude of the resulting pressure surge is closely related to the oscillation period of the cavitation bubble and  its size, which in turn are decided by operating factors (e.g., the initial flow velocity \citep{Bergant1999pipeline,Simpson1991} and pressure differences in the system \citep{Bergant2006Water}). %\zp{in the later stage of our paper, we want to hook back these two things.}

Besides the physical and numerical models of hydraulic systems, the so-called `tube-arrest' method provides an equivalent and low-cost approach to produce transient cavitation bubbles. In a tube-arrest setup, a liquid-filled tube with an initial upward-moving velocity is arrested by a stopper, causing rapid liquid deceleration and the occurrence of cavitation bubbles at the bottom of the tube~\citep{chesterman1952dynamics}.
Based on this technique, \cite{qi2004production} were able to produce bubbles with sizes comparable to the tube diameter, forming a full liquid column separation. 
It was found that the volume of the cavitation bubble could be affected by the initial velocity of the tube. However, the onset criteria of a large cavitation bubble, as well as the dependence of bubble size on operational parameters in a transient process still remains unclear.

In the present work, we proposed a new cavitation number to predict the onset of large cavitation bubbles, and put forward a Rayleigh-type model for the dynamics of a large bubble in a cylindrical tube. The analytical results of bubble size and its oscillation period are essential for estimating the pressure surge in the transient processes. A modified experimental setup based on the `tube-arrest' approach was used to systematically validate the theoretical analysis. 

% In the present work, we designed and manufactured an experimental setup to systematically study the dynamics of large cavitation bubbles based on the tube-arrest principle. We proposed a new cavitation number ($Ca_2$) to predict the onset of large cavitation bubbles, and provide analysis on the dynamics of the cylindrical cavitation bubbles, which is essential in estimating the pressure surge in the transient processes. 

% \zp{may reverse the last two sentences: based on energy conservation, we proposed... then validate with exp ...}%The high speed camera results showed a close relationship between the two phenomenons.

\section{Experimental setup} 
\label{setUpSec}
\begin{figure}
    \centering
    \includegraphics
    [width=4in]{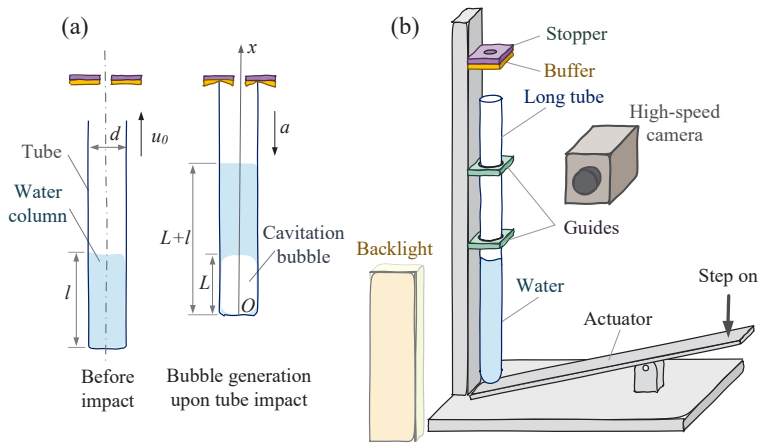}
    \caption{Experimental setup. (a) Cavitation bubble generation process by tube-arrest principle and major parameters (dimensions not to scale), (b) Schematic of test apparatus.}
    \label{fig_exp}
\end{figure}

The modified tube-arrest setup used in the present research is presented in figure~\ref{fig_exp}. 
An acrylic tube (inner diameter $d =16$,~18,~19,~and~20~mm) filled with degassed water to heights of $l = 200$--700 mm is driven upwards by stepping on an actuator until the top of the tube hits a stopper. 
% Through dimensional analysis, it can be derived that the maximum length of the generated cylindrical cavitation bubble $L_{\max} = {d} \cdot f\left( {u_0\sqrt {\frac{\rho }{p}}, \frac{a\rho d}{p}, \frac{l}{d}} \right)$, where $u_0$ and $a$ are the tube impact velocity and acceleration upon impact respectively, and $\rho$ is the density of the liquid. 
According to the Buckingham $\Pi$ theorem of  dimensional analysis and knowledge of influencing factors (such as initial velocity, acceleration and pressure difference) from references in \S\ref{intro}, the maximum length of the generated cavitation bubble can be nondimensionalized by $d$ and is a function of three non-dimensional groups:
\begin{equation}
\label{eq:non-dimensional}
    \frac{L_{\max}}{d} = F(\Pi_1,\Pi_2,\Pi_3)= F\left(\frac{\Delta p}{\rho a l}, {{\frac{\Delta p }{u_0^2 \rho}},  \frac{l}{d}} \right),
\end{equation}
 where $u_0$ and $a$ are the tube impact velocity and acceleration upon impact, respectively. $\rho$ is the density of the liquid, and $\Delta p = p _r - p_v$ is the pressure difference between the reference pressure $p_r$ and the vapor pressure $p_v$. 
Thus, it is intuitive to select $u_0$ and~$a$, in addition to~$l$, as the independent variables for the experiments. In our tests, $u_0$ is controlled by how hard one steps on the actuator. The bottom of the stopper is attached with buffer materials (e.g., rubber and foam) with various combination of stiffness and thicknesses. Taking advantage of the various cushions, $a$ is able to be decoupled from $u_0$ in the tests.

The dynamics of the cavitation bubbles are recorded by a high-speed camera (Phantom~V711, Vision~Research, USA or FASTCAM~Mini UX50, Photron, Japan) with a frame rate of 8,000--10,000~fps. The velocity of the tube during the process is reconstructed from the high-speed videos by tracking a marker on the tube wall, and $u_0$ varies from 1.0 to 6.0 m$\cdot$s$^{-1}$.
The impact acceleration $a$ {\color{black}{(or deceleration in some engineering contexts)}} is measured by an accelerometer (357B03, PCB, USA) attached on the tube sampling at 102,400 Hz, and also evaluated by image processing in some cases (with a discrepancy $<10$\% with accelerometer measurement). The acceleration for all experiments lies in the range from  $-98$ to 23,906~m$\cdot$s$^{-2}$.
The bubble length $L$ is also measured from the calibrated images directly.

The uncertainty of tube inner diameter and liquid column length are less than 1~mm and $10$~mm, respectively.
The impact velocity $u_0$ is obtained by linear regression with 10--15 measurements of the tube displacement, as the motion of the tube is nearly linear right before impact. 
The uncertainty corresponding to the widest 95$\%$ confidence interval of the regression is \SI{0.04}{m\cdot s^{-1}}.
The nominal uncertainty of the acceleration sensor is 2\%.
The measurement error of bubble length obtained from the image is less than 10 pixels (corresponding to $\sim$0.9~mm in physical dimension).

\section{Onset of large cavitation bubbles}
\label{results}
We first address the onset criteria of a cavitation bubble that is big enough to yield liquid column separation. For a cavitation bubble confined in a cylindrical tube, the size of the bubble can be characterized by its maximum length $L_{\max}$ along its axial direction.

\begin{figure}
    \centering
    \includegraphics[width= 0.82 \textwidth]{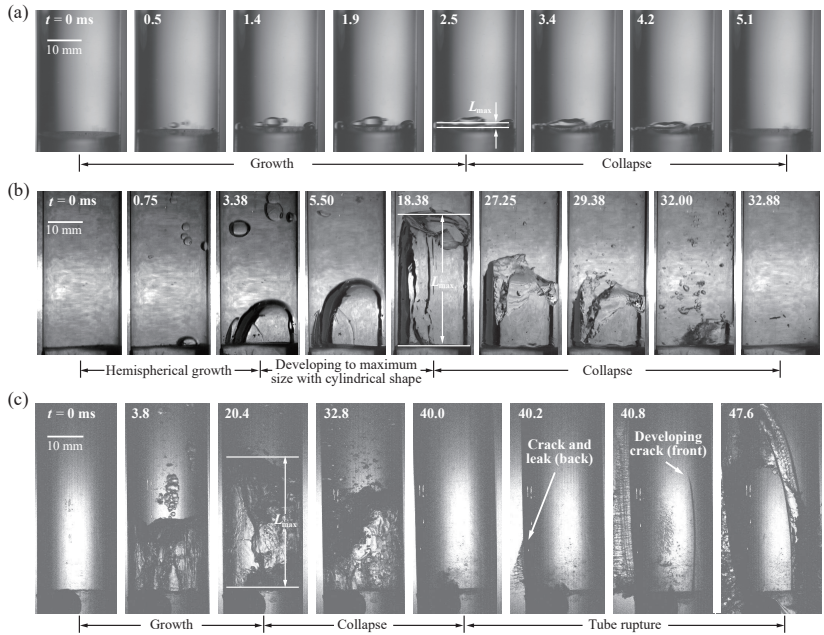}
    \caption{High speed images of typical cases of cavitation bubble(s) formed at the bottom of the tube. $t =0$~ms indicates the moment of tube impacting upon the stopper. (a) Small domed cavitation bubbles. (b) Large cylindrical cavitation bubble. (c) Tube burst caused by the collapse of a large cavitation bubble. $d = 18, 16, 18~\SI{}{mm}$, $l =$~200,~530,~640~\SI{}{mm}, $u_0 =$~1.91,~3.40,~3.08~\SI{}{m\cdot s^{-1}}, and $a =$~843.00,~1857.00,~8050.00~\SI{}{m\cdot s^{-2}}, for case (a), (b), and (c),~respectively.
}
    \label{bubble_example}
\end{figure}

Figure~\ref{bubble_example} shows examples of high speed images of cavitation bubbles generated at the tube bottom. The onset of the bubble occurs immediately after the tube is arrested ($t = $ 0~ms). In figure~\ref{bubble_example}(a), the cavitation bubble grows (0--2.5~ms) to a domed shape and then {\color{black} collapses} (2.5--5.1~ms). The equivalent $L_{\max}$ of the bubble is evaluated by dividing the maximum volume of the bubble by the cross section area of the tube.

Figure~\ref{bubble_example}(b) gives an example of a large cavitation bubble. The bubble grows to a hemisphere until its diameter approaches the inner diameter of the tube (0--3.38~ms). The bubble then undergoes growth along the tube and develops into a large cylindrical bubble to its maximum length $L_ {\max} = $31~mm (at 3.38--18.38~ms). We observe that when $L_{\max}>d$ water column separation is almost guaranteed to happen and we formally address such a cavity `large cavitation bubble' hereafter. After the full water column separation, the first collapse (at 18.38--32.88~ms) happens, and subsequent decaying oscillation lasts until the bubble eventually disappears. The first bubble collapse induces the most violent pressure impulse that scales with water hammer pressure. A rupture event of an acrylic tube was recorded in one of the preliminary tests, where a tube with a thinner wall was used, as shown in figure~\ref{bubble_example}(c) and Supplemental Movie~1. This extreme case demonstrates the power of a large transient cavitation bubble. %Given that the tube inner diameter $d$ is $18$~mm, wall thickness is $2$~mm, and the ultimate strength of the tube material is $\sim$50~MPa, the amplitude of the pressure impulse caused by cavitation collapse can be estimated by the burst pressure of the tube using Barlow's formula, {\color{blue}which is $\sim$$O(10)$~MPa (details in Supplementary Material)}. This estimation agrees with the water hammer pressure {\color{blue}($4.6$~MPa for this particular case)}, as shown in~\S\ref{section:dynamics}.

In our setup, the bulk motion of the water column relative to the tube upon impact can be characterised by the {\color{black}initial velocity and acceleration of the liquid bulk}, which can be approximated by $u_0$ and $a$, respectively. %{\color{magenta} After cavitation onset, the bulk liquid motion relative to the tube can be approximated using the bubble length at the bottom of the tube.}
{\color{black} The justification is as follows.  Right before impact, no bubbles are formed yet and the liquid moves with the tube at $u_0$. Right after the tube is arrested by the stopper in a very short period of time (decelerated at a high $a$), the bulk water column tends to keep moving (at $u_0$) in the tube due to inertia, as if a piston moving in a cylinder. After that, the bulk motion of the water relative to the tube can be estimated using the length of the cavitation bubble (if any appears) by ignoring the boundary layer effect}. The boundary layer thickness at tube side wall $\delta$ can be estimated as $\delta=\sqrt{\nu t}$, where $ t \sim d/u_0$~\citep{Onuki2018Microjet}.
%When the water moves in the tube, the  boundary layer thickness at tube side wall $\delta$ can be estimated as $\delta=\sqrt{\nu t}$, where $ t \sim d/u_0$~\citep{Onuki2018Microjet}. 
Given that $u_0 \sim O(1)$~m$\cdot$s$^{-1}$ and $d \sim O(10)$~mm in our experiments, $\delta / d \sim O(10^{-2})$. 
Thus, the influence of boundary layer on the bulk velocity and acceleration of the water column {\color{black}at tube impact} can be neglected.

\begin{figure}
    \centering
    \includegraphics[width=\textwidth]{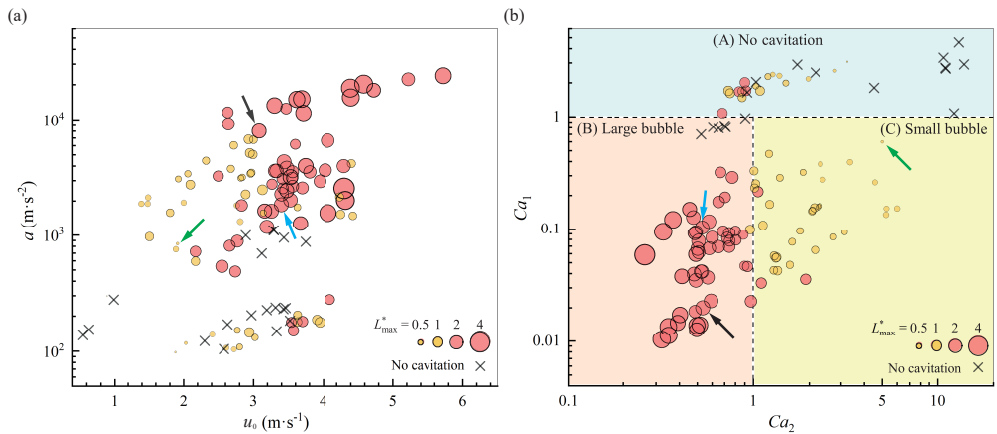}
    \caption{Dimensionless maximum bubble length $L_{\max}^*$ on (a) {\color{black} $u_0$ vs. $a$ and (b) $Ca_2$ vs. $Ca_1$} phase diagrams, respectively. The areas of the filled circles indicate the values of  $L_{\max}^*$, which lie in the range of 0.004--4.25 in the current experiments. For $L_{\max}^*\leq $1 (small bubbles), the circles are filled in yellow. For $L_{\max}^* > 1$, the circles are filled in red indicating large cavitation bubbles. (b) is divided into 3 regimes by $Ca_1$ = 1 and $Ca_2$ = 1. Regime~A ($Ca_1 > $1) represents the no cavitation zone. The sub-domain of $Ca_1 < 1$ indicates cavitation onset, which is separated into two regimes by $Ca_2 = 1$: onset of large cavitation bubbles (regime~B, $Ca_2 < 1$) and small cavitation bubbles (regime~C, $Ca_2 \geq 1$), respectively. The particular cases illustrated in figure~\ref{bubble_example}(a), (b), and (c) are marked with green, blue and black arrowheads, respectively.
}
    \label{fig_Ca12}
\end{figure}

Figure~\ref{fig_Ca12} summarizes the experimental results of dimensionless maximum bubble length $L _ {\max} ^* = L _ {\max} / d$. The results are first presented in a dimensional {\color{black} $u_0$ vs. $a$} diagram as shown in figure~\ref{fig_Ca12}(a). No distinct regime of large cavitation bubbles with $L_{\max}^* > 1$ distributed can be found on this diagram. Thus, the large cavitation bubble onset cannot be predicted directly using either $u_0$ or $a$.  Instead, we show here that onset criteria of large cavitation bubbles can be represented with a phase diagram in terms of two non-dimensional parameters. 

The first parameter is a cavitation number $Ca_1$ proposed by \cite{Pan2017Cavitation}:
\begin{equation}
Ca_1 = \frac{ p_r - p_v }{\rho a l} \approx \frac{p_\infty}{\rho a l},
  \label{eq:ca1}
\end{equation}
where $p_r$ is the reference pressure and $p_v$ is the saturated vapor pressure of the liquid, respectively. In the current experiments, $p_\infty \approx p_r \gg p_v$ is approximately the atmosphere pressure. {\color{black} The static pressure of the water column, which is one order of magnitude smaller than $p_{\infty}$, is neglected}. This cavitation number applies to transient scenarios where the influence of liquid acceleration on pressure variation is much greater than that of the flow velocity, which is the case in our experiments. By this criterion, $Ca_1<1$ predicts cavitation onset.

We now investigate the size of the cavitation bubbles. Before the collision, the kinetic energy of the water column is at its maximum, which is $E_{k_{\max}}=0.5\rho {\pi}l(d/2)^2  u_0^2$. During the collision, the tube is arrested in a very short time while the liquid keeps the motion upwards, forming the cavitation bubble(s) at the bottom of the tube. The bubble(s) then grow (and merge) to a maximum length until the bulk velocity of the water column vanishes. Given that $p_ \infty \gg p_v$, the potential energy of the bubble is approximately $E_{p_{\max}}  = p_\infty {\pi}(d/2)^2 L_{\max}$ at its maximum length. 

It is assumed that the energy dissipation due to pressure waves and viscosity is negligible. Balancing $E_{k_{\max}}$ and $E_{p_{\max}}$ gives $L_{\max}/l=0.5\rho u_0^2/p_{\infty}$; and further rearrangement leads to:
\begin{equation}
\label{Ca_2}
Ca_2=\frac{d}{L_{\max}}=\left(\frac{d}{l}\right)\left( \frac{p_\infty}{0.5\rho u_0^2}\right) = \frac{1}{l^*}Ca_0,
\end{equation}
where $l^* = l/d$ is the non-dimensional length of the water column indicating the slenderness of the water column, and $Ca_0=p_\infty/0.5\rho u_0^2$ is the classic cavitation number.  $Ca_2$ is a new cavitation number that can be considered as a modified version of the classic cavitation number. $Ca_2$ is the second non-dimensional number to characterize the onset of the large cavitation bubble. 

\begin{figure}
    \centering
    \includegraphics[width=3.0in]{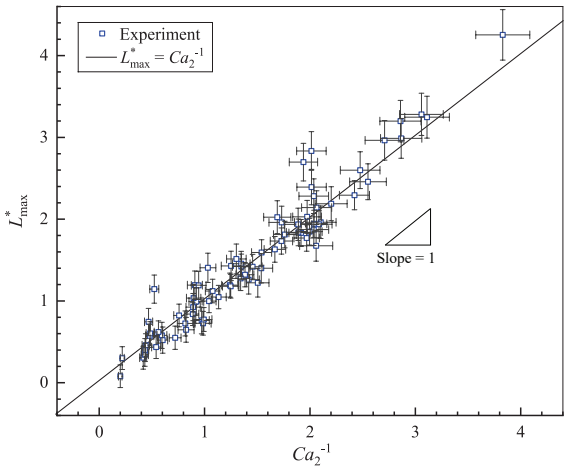}
    \caption{Experimental validation of Eq.~\eqref{eq:L*=1/Ca2},  $L_{\max}^* = Ca_2^{-1}$.
}
    \label{fig_Ca2_L}
\end{figure}

Rearrangement of Eq.~\eqref{Ca_2} yields:
\begin{equation}
    \label{eq:L*=1/Ca2}
    L_{\max}^*=Ca_2^{-1}, 
    \end{equation}
meaning that $Ca_2$ (to be more specific, $Ca_2^{-1}$) is a direct measure of the non-dimensional maximum bubble size (i.e., the maximum cavitation length $L_{\max}$ relative to the tube diameter~$d$). 
This theoretical dependence of $L_{\max}^*$ on the new cavitation number $Ca_2$ by Eq.~\eqref{eq:L*=1/Ca2} is validated, and agrees well with the experimental results, as shown in figure~\ref{fig_Ca2_L}. 

The onset of large cavitation bubbles requires both $Ca_1<1$ (onset threshold for cavitation, small or large) and $Ca_2 < 1$ (threshold for large bubbles) simultaneously. Note that $Ca_1 = \Pi_1$ and $Ca _2 = 2 \Pi_2 / \Pi_3$. Thus $L_{\max}^* = \Pi_1 ^0 (2 \Pi_2/ \Pi_3 ) ^{-1}$.
% This indicates that in the context of, for example, a hydraulic system, severe water column separation occurs when the valve shuts down too fast (corresponding to small $Ca _1$) while the flow goes too fast through a conduit that is too long (corresponding to small $Ca _2$).
 This indicates that in the context of, for example, a hydraulic system, severe water column separation occurs when {\color{black}(i) the valve shuts down too fast while (ii) the fluid flows too fast through a conduit that is too long. Shutting too fast corresponds to a small $Ca_1$: a pressure drop due to an inertial force induced by the high acceleration exceeds the tensile strength of the liquid and cavitation bubbles appear. Fluid flowing too fast through a long conduit corresponds to a small $Ca_2$: the kinetic energy associated with a long and fast-moving liquid column is high enough to provide the potential energy needed to generate a large bubble.}

These comprehensive criteria are then validated against the experimental data on the {\color{black} $Ca_1$ vs. $Ca_2$} diagram, as shown in figure~\ref{fig_Ca12}(b). 
The whole domain is divided into three regimes by $Ca_1=1$ and $Ca_2=1$.
As expected, almost all the no-cavitation events are located in regime~A ($Ca_1>1$, blue area in figure~\ref{fig_Ca12}); while when $Ca_1<1$ (red and yellow areas in figure~\ref{fig_Ca12}, cavitation bubbles are consistently observed.
These results agree with the finding in \cite{Pan2017Cavitation}.
The sub-domain of $Ca_1<1$ is further split into two regimes by $Ca_2=1$.
Regime~B ($Ca_2<1$, red area in figure~\ref{fig_Ca12}) represents the criterion of large cavitation bubble onset, and regime~C ($Ca_2\geq1$, yellow area in figure~\ref{fig_Ca12}) is for small cavitation bubbles, respectively. Again, experimental data fall in the corresponding regimes and agree well with the theoretical prediction. 
Typical high-speed videos with no cavitation, small and large cavitation bubbles can be found in the Supplemental Materials. 

\section{Dynamics of cylindrical cavitation bubbles}\label{section:dynamics}

We next investigate the dynamic behavior of a large cavitation bubble. 
As described above (see also figure~\ref{bubble_example}), a large cavitation bubble first grows radially until its interface reaches the inner wall of the tube. 
It then develops into a cylindrical bubble due to the confinement of the tube. 
Over the majority of the bubble life cycle, the cavitation bubble grows and collapses in one dimension along its axis of symmetry. This feature resembles the dynamics of bubbles in a tube with smaller scales as in \cite{ory2000growth}, \cite{yin2004bubble}, and \cite{sun2009growth}. Based on this observation, we ignore the initial hemispherical growth of the bubble, and develop a Rayleigh-type equation for the dynamics of cylindrical cavitation bubbles as follows.
\begin{figure}
    \centering
    \includegraphics[width=3.2in]{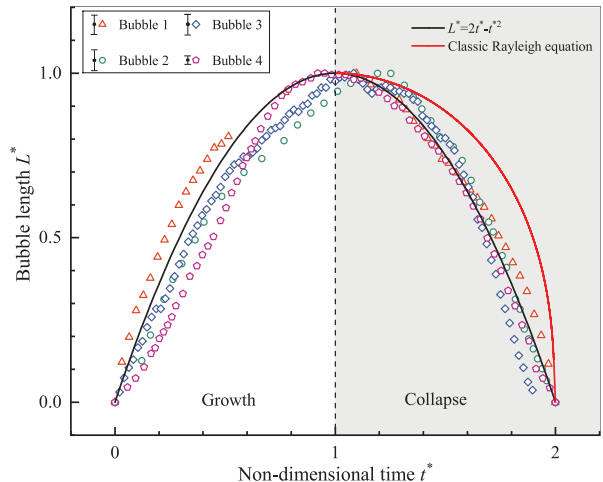}
    \caption{Evolution of normalized length of cylindrical cavitation bubbles $L ^* = L / L_{\max} $. $t^* = t/ T_c$. The uncertainty of the non-dimensionalized bubble length for each case is shown next to the corresponding legend. Both experimental and theoretical results show that the growth and collapse of the bubbles are almost symmetrical over time. Compared with the \textit{infinite} collapsing speed of a spherical cavitation bubble by the Rayleigh solution (red line), a cylindrical cavitation bubble has a \textit{finite} collapsing velocity instead (black line).
 }
    \label{fig_Lt}
\end{figure}

\begin{figure}
    \centering
    \includegraphics[width=3in]{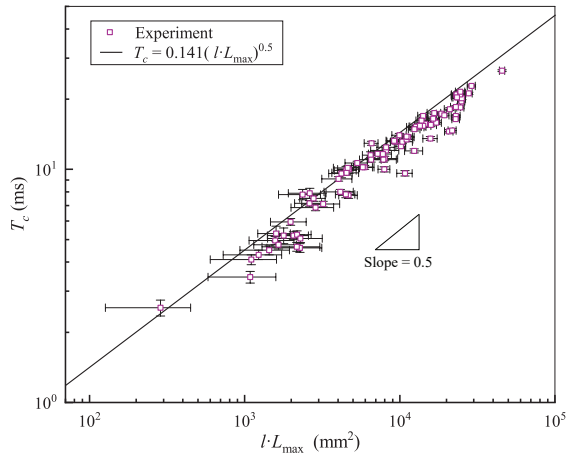}
    \caption{Dependence of collapse time $T_c$ (from maximum length to first collapse) of cylindrical cavitation bubbles on $l L_{\max}$. 
}
    \label{fig_TL}
\end{figure}

As shown in figure~\ref{fig_exp}, since the side boundary of the cylindrical bubble is close to the inner wall of the tube, the velocity of the liquid $u(x,t)$ can be simplified as $u(t)$. 
Ignoring the mass diffusion between the gas and liquid phases, the continuity equation of the liquid reads $u(t)=\mathrm{d} L/\mathrm{d} t$.
Further ignoring the viscosity of the liquid, the Euler equation for the motion of the fluid in the $x$~direction is
\begin{equation}
    \rho \left(\frac{\partial u}{\partial t}+u\frac{\partial u}{\partial x}\right)=-\frac{\partial p}{\partial x}.
    \label{NS_x}
\end{equation}
Substituting  the continuity equation into Eq.~\eqref{NS_x} leads to
\begin{equation}
    \frac{\mathrm{d}^2 L}{\mathrm{d} t^2}=-\frac{1}{\rho } \frac{\partial p}{\partial x},
    \label{boundary motion}
\end{equation}
where $\mathrm{d}^2L/\mathrm{d}t^2$ is the acceleration of the vapor-liquid interface. Integrating Eq.~\eqref{boundary motion} from the top of the bubble $x=L$ to the free surface $x=L+l$, and considering pressure at the bubble boundary $p_{L}\approx p_v \ll p_{L+l}$, where $p_{L+l}$ is the pressure at the free surface, the acceleration of cavitation bubble boundary can be simplified as
\begin{equation}
    {\frac{\mathrm{d}^2L}{\mathrm{d}t^2}}= - \frac{p_{L+l}}{\rho l}.
   \label{motion_result_1}
\end{equation}

Assuming the tube collides on the stopper at $t=0$, we then have $\mathrm{d}L/\mathrm{d}t|_{t=0}=u_0$ and $L|_{t=0}=0$ as initial conditions. Integrating Eq.~\eqref{motion_result_1} twice gives the equation for the dynamics of the cylindrical cavitation bubbles during its growth and first collapse: $ L=u_0 t{\color{black}+}\frac{1}{2}\frac{\mathrm{d}^2L}{\mathrm{d}t^2} t^2$. Further non-dimensionalization leads to
\begin{equation}
\label{eq_L*t*}
   L^*=2t^*-t^{*2},
\end{equation}
where $t^*=t/T_c$ and $L^*=L/L_{\max}$. $T_c$ is the collapse time  (duration from maximum length to collapse) of the bubble. The time interval between bubble onset and its first collapse is then $2T_c$. This parabolic form of bubble length over time is experimentally validated as shown in figure~\ref{fig_Lt}. The experimental results agree well with Eq.~\eqref{eq_L*t*}. Compared to the infinite collapsing speed of a spherical bubble by the Rayleigh solution, {\color{black} the collapse of a cylindrical bubble is `milder' and} has a finite collapse velocity $u_0$ instead, which is the same as the initial tube collision speed. 

Given this knowledge, we can evaluate the water hammer pressure in the case of figure~\ref{bubble_example}(c) as $\rho c u_0$, where $c$ is the speed of the sound of water, and $u_0=$~\SI{3.08}{m\cdot s^{-1}} in the particular case. we can readily estimate the pressure impulse generated by the first collapse of the cylindrical bubble in a tube during a transient process,  which is of great importance in future engineering practice such as sudden valve closure in pipelines. Considering that $\rho \approx$~1,000~$\rm{kg / m^3}$ and $p_ {\infty} \approx 100,000~\rm{Pa}$ for water under atmosphere pressure, the collapse time of a cylindrical cavitation bubble can be estimated as
\begin{equation}
\label{Tc-2}
  T _ c=\sqrt{2}\sqrt{lL_{\max}}\sqrt{\frac{\rho}{p_\infty}}\approx0.141\sqrt{lL_{\max}} = 0.141\sqrt{\frac{ld}{Ca_2}}.
\end{equation}
% \begin{equation}
% \label{Tc-2}
%   T _ c=1.414\sqrt{lL_{\max}}\sqrt{\frac{\rho}{p_\infty}}\approx0.141\sqrt{lL_{\max}} = 0.141\sqrt{\frac{ld}{Ca_2}}.
% \end{equation}
Validation of Eq. \eqref{Tc-2} is shown in figure~\ref{fig_TL}, where $T_c$ of cylindrical bubbles are well predicted. Note that for a spherical cavitation bubble in an infinite domain, the Rayleigh collapse time is $T_{c,R}=0.915R_{\max}\sqrt{\rho /p_\infty}\approx0.091R_{\max}$, while for a cylindrical cavitation bubble $T_c$ depends on both the length of the water column $l$ and the maximum bubble length $L_{\max}$. Interestingly, {\color{black} the length scale ($\sqrt{lL_{\max}}$)} in Eq.~\eqref{Tc-2} is the geometric average of $L_{\max}$ and $l$, which is an analogy to the maximum bubble radius in the classic Rayleigh time $T_{c,R}$.  {\color{black} 
It is also interesting to note that $T_c > T_{c,R}$ for bubbles with the same length scale ($\sqrt{lL_{\max}}=R_{\max}$), indicating a slower (and milder)  collapse of a cylindrical bubble compared to a spherical one.}
Equation \eqref{Tc-2} is useful in engineering practice in order to  estimate the bubble oscillation period when large cavitation bubbles are present.

\section{Conclusions} 
In the current paper, we investigate the dynamics of a large cavitation bubble confined in a tube during a transient process by the `tube-arrest' technique. We focus on the onset criteria of a large bubble that can cause full water column separation, as well as the maximum length and collapse time of the cavitation bubble.

By considering energy conservation, we find that the length of the cavitation bubble in the tube can be characterized by a non-dimensional parameter $Ca_2=l^*{^{-1}}Ca_0$, which is a modified version of the the classic cavitation number $Ca_0$. When $Ca_2<1$, large cavitation bubbles are developed in the tube.  
In addition, we recall the onset criteria~($Ca_1<1$) of cavitation bubbles during a transient process proposed by \cite{Pan2017Cavitation}.
Thus, the large cavitation bubbles can only occur when both $Ca_1<1$ and~$Ca_2 <1$ are simultaneously satisfied. Our findings indicate that once cavitation occurs, the maximum size of cavitation bubbles is only determined by $Ca_2$ and is independent to $Ca_1$, meaning that the severity of the transient process does not affect the severity of the cavitation damage due to large bubbles. 
These comprehensive criteria for large cavitation bubbles are validated by systematic experiments based on the tube-arrest approach. 
% The tube-arrest setup takes advantage of the buffer and can decouple the  velocity and the acceleration of the liquid in the tube, thus decouple $Ca_1$ and $Ca_2$.  
%With different combinations of parameters including liquid column length, collision velocity and acceleration, we have obtained no cavitation, large cavitation bubbles and small cavitation bubbles respectively. These experiment results are well organized by two non-dimensional number $Ca_1$ and $Ca_2$, where $Ca_1<1$ is the condition for cavitation onset and $Ca_2$ could further predict the maximum size of cavitation bubbles when cavitation occurs. Remarkably, the effect of $Ca_2$ on the maximum size of cavitation bubbles is independent, that is, once cavitation occurs, the maximum size of cavitation bubbles is only determined by $Ca_2$ and is almost unaffected by $Ca_1$. The definition of $Ca_2$ indicates that a longer length of water column with higher initial collision speed is more likely to produce large bubbles.

We also propose a Rayleigh-type equation for the dynamics of cylindrical cavitation bubbles and predicted the growth and collapse process over time. Contrary to the infinite collapse velocity of a spherical cavitation bubble as predicted by the classic Rayleigh equation, a cylindrical bubble has a finite collapse velocity equivalent to the initial impact velocity (or the speed of the liquid relative to the pipe). It is also revealed that the collapse time of the cylindrical bubble depends on not only the maximum bubble length, but also the water column length. 
% The theoretical results are also validated by experiments.

The results of this research may provide guidelines in the design and operation of hydraulic systems considering safety during transient processes. It should be noted that in addition to previous knowledge on transient cavitation onset, the conditions for large bubble generation, as well as the size and collapse time of large cylindrical cavitation bubbles are mainly determined by the initial steady-state flow conditions. This finding can be applied to alleviate the damage to the hydraulic conduits from extreme cavitation that forms water column separation.

%We also proposed a 1D model for the dynamics of the produced cylindrical cavitation bubbles. The growth rules of the cavitation bubbles can be predicted by a normalized quadratic equation. Compared to the Rayleigh solution for a spherical bubble, the cylindrical bubble has a finite collapse velocity equivalent to the impact velocity and a collapsing time of $0.141\sqrt{lL_{max}}$.

%The findings of this paper may help us to control the cavitation bubbles size during the transient process, thus avoiding the superposition of water hammer pressure and cavitation bubble collapse pressure to the harm of pipeline system by controlling the duration time of large cavitation bubbles.

%In this paper, we propose to find a way to produce large size cavitation bubbles, figure out the factors which affect the cavitation bubble size and use a dimensionless number to represent the bubble size. In addition to the maximum bubble size, the dynamic growth of cylindrical cavitation bubbles is also taken into account by extending the Rayleigh$-$Plesset equation\c. experimental results of relationship between bubble size and cavitation number $Ca_2$ with the theory, which shows the results agree well (figure~\ref{Ca2}. And the changing rule of lifecycle time of the bubbble with its maximum length from experiments also consistent with formula~\eqref{motion result} we have put up (figure~\ref{111}).

\section*{Acknowledgments} 
We gratefully acknowledge inspiring discussions with H. A. Stone, D. Lohse and Y. Tagawa, and the technical supports by H. Li, D. Wang and Y. Guo
and the help from our research group. The work was partially supported by National Natural Science Foundation of China (no.~52076120 and no.~11861131005), the Skate Key Laboratory of Hydroscience and Engineering (sklhse-2019-E-02), and Tsinghua University-University of Waterloo Joint-Research Centre.

\bibliographystyle{jfm}
\bibliography{CavitationLibrary}

\end{document}